\documentclass{article}
\usepackage{amssymb}

\usepackage{amsmath}

\newenvironment{proof}[1][Proof]{\textbf{#1.} }{\ \rule{0.5em}{0.5em}}
\input{tcilatex}

\begin{document}

\title{Gauge theories on noncommutative euclidean spaces}

\author{Albert Schwarz \\
Department of Mathematics, University of California, Davis, CA 95616}
\maketitle

\begin{abstract}
We consider gauge theories on noncommutative euclidean space . In
particular, we discuss the structure of gauge group following standard
mathematical definitions and using the ideas of hep-th/0102182 .
\end{abstract}

\bigskip\ 

The goal of this note is to consider gauge theories on noncommutative
euclidean space and, in particular, to study the structure of gauge group.
This group was analyzed by J.Harvey in recent paper [1]. It was suggested in
this paper that the definition of the gauge group ''presumably can be
derived from the first principles''. We would like to analyze the relation
of Harvey's definition to the standard mathematical definition using as
a starting point some ideas of [2], in particular, the idea that the theory
becomes more transparent if along with simple modules $A^{n}$ we consider
more complicated modules $\mathcal{F}_{rn}$. (The central point of [2]-the
suggestion to work with unitized algebras-is mentioned only in passing at
the very end.)

Mathematical definition of a gauge field is based on a notion of connection
on a module $E$ over associative algebra $A$. There exist different versions
of this notion (see [4] for details, [5] for more general treatment). Our
consideration does not depend on these subtleties. We can use, for example,
the very first definition [3]; in this definition linear operators $\nabla
_{1,...,}\nabla _{d}$ specify a connection on right $A$-module $E$ if they
satisfy Leibniz rule: 
\begin{equation*}
\nabla _{\alpha }(ea)=(\nabla _{\alpha }e)\cdot a+e\partial _{\alpha }a
\end{equation*}%
where $\partial _{1},...,\partial _{d}$ are derivations on $A$, $e\in E$, $%
a\in A$. One assumes that these derivations (i. e. infinitesimal
automorphisms) constitute a basis of a Lie algebra. By definition a gauge
field is a unitary connection (i. e. $\nabla _{\alpha }$ should be anti
Hermitian operators).

It is supposed usually that $A$ is a unital Banach algebra over $\mathbb{C}$
and  $E$ is a Hilbert $A$-module (i. e. $E$ is equipped with $A$%
-valued Hermitian inner product \TEXTsymbol{<} , \TEXTsymbol{>}; then the
condition of unitarity of connection takes the form

\begin{equation*}
<\nabla _{\alpha }a,b>+<a,\nabla _{\alpha }b>=\partial _{\alpha }<a,b>
\end{equation*}

(By definition a Banach algebra is an associative algebra with a norm and an
involution obeying natural conditions. The vector space $A^{n}$ of column
vectors with entries from $A$, considered as a right $A$-module, can be
equipped with $A$-valued Hermitian inner product $%
a_{1}^{+}b_{1}+...+a_{n}^{+}b_{n}$. Hence all projective $A$-modules, i. e.
direct summands in $A^{n}$, can be considered as Hilbert modules.)

For every unital Banach algebra $A$ we can construct a group $U(A)$
consisting of unitary elements of $A$. The topology of this group and of
groups $U_{n}(A)=U($Mat$_{n}(A))$ is closely related to the $K$-theory of $A$%
; namely 
\begin{equation*}
K_{i}(A)=\pi _{i-1}(U_{\infty }(A))
\end{equation*}%
where $U_{\infty }(A)=\cup U_{n}(A)$ is a union (or more precisely direct
limit) of groups $U_{n}(A)$. (This is one of possible definitions of $K$%
-groups.) Notice, that $\pi _{i}(U_{\infty }(A))=\pi _{i+2}(U_{\infty }(A))$
(Bott periodicity theorem) for $i\geq 0;$ using this periodicity we can
define groups $\pi _{i}(U_{\infty }(A))$ for negative $i$.

The definitions and results that we formulated for Banach algebras can be
applied also to more general  algebras, equipped with antilinear 
involution.
  
In the definition of connection we have used the role of gauge
transformations is played by unitary endomorphisms of $E$ (unitary $A$%
-linear maps of $E$ into itself). This follows from the fact that for every
unitary endomorphism $\varphi $ the correspondence $\nabla _{\alpha
}\rightarrow \nabla ^{\prime }=\varphi \nabla _{\alpha }\varphi ^{-1}$
transforms a unitary connection $\nabla _{\alpha }$ into a unitary
connection $\nabla ^{\prime }$; corresponding curvatures are related in
standard way: $F_{\alpha \beta }^{\prime }=\varphi F_{\alpha \beta }\varphi
^{-1}$. Considering the curvature of unitary connection as a field strength
of a gauge field we obtain that all reasonable action functionals are
invariant with respect to unitary endomorphisms. It is possible to consider
the group of unitary endomorphisms $U($End$_{A}E)$ (the group of unitary
elements of endomorphism algebra End$_{A}E$) as a gauge group. If $E=A^{1}$ (%
$U(1)$-gauge field in the terminology of physicists) this group is
isomorphic to $U(A)$ and in the case of $E=A^{n}$ (the case of $U(n)$-gauge
fields), it is isomorphic to $U_{n}(A)=U($Mat$_{n}(A))$. (We use the
notation Mat$_{n}(A)$ for the algebra of $n\times n$ matrices with entries
from $A$. The endomorphisms of $A^{n},$ considered as a right module, can be
identified with elements of Mat$_{n}(A)$ acting on $A^{n}$ from the left.)
It is important to notice that usually physicists work with a little bit
different definition of gauge field, that is equivalent to the above
definition restricted to the modules $E=A^{n}$ (free modules).

The above consideration can be applied also to non-unital algebras; the only
essential difference is that the elements of Mat$_{n}(A)$ don't exhaust in
this case all endomorphisms of $A^{n}$(the modules $A^{n}$ cannot be
considered as free modules). The algebra $M(A)$ of endomorphisms of $A^{1}$
is called multiplier algebra of $A;$ using this definition one can represent
endomorphisms of $A^{n}$ as matrices with entries from $M(A)$. We see that $%
U($End$_{A}A^{n})=U($Mat$_{n}(M(A)))$.

As we have seen one can consider $U($End$_{A}E)$ as a gauge group. This
means that we can identify two gauge fields connected by gauge
transformation and use the integration over the quotient space $\mathcal{C}$ 
$/G$ where $G=U($End$_{A}E)$ to quantize the gauge theory. (Here $\mathcal{C}
$ stands for the space of all gauge fields.) However, sometimes it is more
convenient to introduce a stronger notion of gauge equivalence replacing $%
G=U($End$_{A}E)$ with its subgroup $G^{\prime }\subset G$. As long as $%
G/G^{\prime }$ is compact we can use $\mathcal{C}$ $/G^{\prime }$ instead of 
$\mathcal{C}$ $/G$ in the quantization procedure. However, if $G^{\prime
}$ acts freely on $\mathcal{C}$ it is more convenient to work with $\mathcal{%
C}$ $/G^{\prime }$.(This space is non-singular and its homotopy groups can
be calculated in terms of homotopy groups of $G^{\prime }$.)

For example, if we work with ordinary $U(n)$-gauge theory on compact
manifold $X$ (i. e. $A=C(X)$ is a commutative algebra of functions on $X$)
and $E=A^{n}$ is a trivial Hermitian vector bundle then $G=U$(End$_{A}E$)
consists of functions on $X$ taking values in unitary matrices. In this case
it is more convenient to consider as a gauge group the subgroup $G^{\prime }$
of $G$ consisting of functions $\varphi \in G$ obeying $\varphi (x_{0})=1$
for a fixed point $x_{0}$ of $X$. The quotient group $G/G^{\prime }$ can be
identified with the group of global gauge transformations. Notice that the
separation of gauge transformations (that we use to define gauge classes)
and global symmetry transformations is not physical in this case (it depends
on the choice of $x_{0}$.) However, sometimes the reduction of $G$ to $%
G^{\prime }$ is prompted not only by mathematical convenience, but also by
physical considerations. (It can happen, that it is necessary to consider
observables that are $G^{\prime }$-invariant, but not $G$-invariant.)

In particular, if we consider ordinary $U(n)$-gauge theory on $\mathbb{R}%
^{d} $, the action functional is invariant with respect to gauge
transformations corresponding to $U(n)$-valued functions $g(x)$, that have a
limit as $x\rightarrow \infty $. However, it is reasonable to consider as a
gauge group a smaller group $G^{\prime }$ imposing a condition $%
\lim_{x\rightarrow \infty }g(x)=1$. (This means that we exclude global gauge
transformations.)

The idea that one can modify the notion of gauge group without changing
physics is reinforced by obvious remark that we can impose conditions
partially removing gauge freedom (i. e. making the gauge group smaller).

Let consider now gauge theories on noncommutative euclidean space $\mathbb{R}%
^{d}$ following the ideas of [2].

In this case we can work with various algebras corresponding to different
behavior of gauge fields at infinity. Let us start with a (non-unital)
algebra $A=$ $S(\mathbb{R}_{\theta }^{d})$ of Schwartz functions on $\mathbb{%
R}^{d}$ equipped with star-product. Every connection on $A^{n}$ has the form 
\begin{equation}
\nabla _{\mu }=\partial _{\mu }+a_{\mu }
\end{equation}%
where $a_{\mu }$ is an $n\times n$ matrix with entries the multiplier
algebra $M(A)=M($ $S(\mathbb{R}_{\theta }^{d}))$.One can consider the 
multiplier algebra as an algebra of generalized functions ( distributions) 
with the multiplication  defined as a
star-product.
In the case $\theta =0$ the 
algebra $M(A)\;$
consists of  smooth functions, having derivatives with at
most polynomial growth at infinity. For nonvanishing $\theta$ the 
description of $M(A)\;$ is more complicated; see [11],[12],[13].
This algebra essentially depends on the choice of $\theta$. 
 One can prove  that for 
nondegenerate $\theta$  a  continuous functional on Schwartz space
$S(\mathbb{R})^d$ ( a  distribution) specifies an element of
$M(A)=M($ $S(\mathbb{R}_{\theta }^{d}))$ if at infinity ( i.e.in the 
complement to a compact set) it can be represented by a smooth function
with all derivatives bounded by polynomials of the same degree  [13].
However,there exist multipliers that do not belong to this class [13].
 
Notice, that we worked with right modules; then multipliers are 
realized by means of multiplication from the left. In the case of left
modules multipliers act from the right. There exists an interesting
algebra consisting of distributions that can be considered as 
left and 
right multipliers at the same time [11],[12],[13].

Considering $a_{\mu }$ as a gauge field we can say that our
gauge fields not necessarily decrease at infinity (they can even grow, but
at most polynomially). Therefore we should impose an additional condition
that gauge fields at hand have finite euclidean action (or finite energy).
In the ideology of functional integral this condition follows from the fact
the contribution of fields with infinite action vanishes. Notice that
instead of the algebra $A=$ $S(\mathbb{R}_{\theta }^{d})$ we can work, for
example, with algebra of functions on $\mathbb{R}^{d}$ that have derivatives
of all orders and all these derivatives tend to zero at infinity (the
multiplication is again defined as a star-product). This algebra is bigger,
it has less connections (gauge fields are bounded in this case). However,
this makes no difference: gauge fields having finite action are the same
(if we impose some mild regularity condititions).

The condition of finiteness is a complicated non-linear condition.In the
case when the dimension of the space is at least 4  we
replace it with a condition that the gauge field is gauge trivial at
infinity. We say that a connection on $A^{n}=(S(\mathbb{R}_{\theta
}^{d}))^{n}$ is gauge trivial at infinity if it can be represented in the
form%
\begin{equation}
\nabla _{\mu }=T\circ \partial _{\mu }\circ S+(1-TS)\circ \partial _{\mu
}+\sigma _{\mu }
\end{equation}%
where $T,S\in $End$_{A}(A^{n})$ are operators obeying $1-TS=\Pi \in
A,1-ST=\Pi ^{\prime }\in A$ and $\sigma _{\mu }$ is small at infinity in
appropriate sense. We will always assume that $T$ belongs to $H\Gamma
_{1}^{m,m_{0}},$ then there exists such $S\in H\Gamma _{1}^{-m,-m_{0}}$ that 
$1-TS\in A$, $1-ST\in A$. (See [6],[2] or Appendix for the definition of the
class of hypoelliptic symbols $H\Gamma _{\rho }^{m,m_{0}}$ and for the
definition of the class $\Gamma _{\rho }^{m}$; roughly speaking $a\in \Gamma
_{\rho }^{m}$ if $\left\| a\right\| \leq $ const $\left\| x\right\| ^{m}$ at
infinity and $T\in H\Gamma _{\rho }^{m,m_{0}}$ if at infinity $T\in \Gamma
_{\rho }^{m}$ and $T^{-1}\in \Gamma _{\rho }^{-m_{0}}$.) We will make
precise the statement that $\sigma _{\mu }$ is small at infinity requiring
that $\sigma _{\mu }\in \Gamma $ where $\Gamma $ stands for the union of $%
\Gamma _{1}^{m}$ with $m<-1$ (this means that $\sigma _{\mu }$ tends to zero
faster than $\Vert x\Vert ^{-1},$ i.e. faster than the first term in (2) ).
Let us denote by $\mathcal{C}(T)$ the class of connections that can be
represented in the form (2)\ with fixed $T$ and $\sigma _{\mu }\in \Gamma $.
Considering $T\in H\Gamma _{1}^{m,m_{0}}$ as a matrix valued function on $%
\mathbb{R}^{d}$ we obtain an element of $\pi _{d-1}(GL(n))=\pi _{d-1}(U(n))$
as a homotopy class of the map of large sphere $S^{d-1}\subset \mathbb{R}%
^{d}$.The class $H\Gamma _{1}^{m,m_{0}}$ consists of components labelled
by
elements of $\pi _{d-1}(U(n))$. Let us fix an element $T_{k}$ in every
component of $H\Gamma _{1}^{0,0}$ and define $\mathcal{C}^{(k)}$ as $%
\mathcal{C}(T_{k})$.

If $T$ and $T_{k}$ determine the same element of $\pi _{d-1}(U(n))$ one can
prove that a gauge field (unitary connection) belonging to $\mathcal{C}(T)$
is gauge equivalent to a gauge field from $\mathcal{C}^{(k)}=\mathcal{C}%
(T_{k})$ (i.e. these two gauge fields are related by unitary endomorphism).\
Moreover, in the case $\theta \neq 0$ this statement remains correct if we
impose weaker condition that $T$ and $T_{k}$ determine the same element of
stable homotopy group $\pi _{d-1}(U(\infty ))$ ( we will prove this
statement below for the case of nondegenerate $\theta $ ).

Working with gauge fields (unitary connections) it is convenient to replace
(2) by an explicitly unitary expression 
\begin{equation}
\nabla _{\mu }=T\circ \partial _{\mu }\circ T^{+}+\Pi \circ \partial _{\mu
}\circ \Pi +\rho _{\mu }
\end{equation}%
where $T\in H\Gamma _{1}^{0,0}$,
 $\Pi =1-TT^{+}\in A$, $\Pi ^{\prime
}=1-T^{+}T\in A$, $\rho _{\mu }=\rho _{\mu }^{+}\in \Gamma $. It is easy to
check that every unitary connection belonging to $\mathcal{C}(T)$ where $%
T\in H\Gamma _{1}^{0,0}$ can be represented in the form (3). Notice, that in
the case of non$\deg $enerate $\theta $ we can consider elements of Mat$%
_{n}(M(A))$ as pseudodifferential operators acting on $S(\mathbb{R}^{m})$
where $d=2p$. The topological class of $T$ can be considered as index of
corresponding pseudodifferential operator $\widehat{T}$. Without loss of
generality we can assume that either $T^{+}T=1$ (i.e. Ker $\widehat{T}=0$)
or $TT^{+}=1$ (i.e. Ker $\widehat{T^{+}}=0$).

We conjecture that calculating correlation functions we can do functional 
integral
over fields that are gauge trivial at infinity. In very vague way one can
say that ``almost all'' (but not necessarily all) fields having finite
action are gauge trivial at infinity.(One can say that the finiteness of 
action implies gauge triviality  for fields obeying some regularity 
conditions at infinity.Some results of this kind can be derived in
commutative case from Uhlenbeck theorem [10])

It is convenient to modify the definition of gauge triviality at infinity in
the following way: we say that the gauge field belongs to the class $%
\mathcal{C}_{m}$ if $T$ in (3) belongs to $H\Gamma _{1}^{0,0}$ and $\rho
_{\mu }$ belongs to $\Gamma ^{m}$ where $\Gamma ^{m}$stands for the union of 
$\Gamma _{1}^{m^{\prime }}$with $m^{\prime }<m$. We always assume
that $m\leq -1$; under this assumption $\mathcal{C}_{m}\subset \mathcal{C}%
_{-1},$ i.e. a gauge field of the class $\mathcal{C}_{m}$ is gauge trivial
at infinity. It easy to check that the euclidean action of a gauge field
from $\mathcal{C}_{m}$ is finite if $m\leq 1-\frac{d}{2}$. (Here $d$ stands
for the dimension of noncommutative euclidean space.)

Let us consider for definiteness the case $d=4$. Noticing that $\pi
_{3}(U(n))=\mathbb{Z}$ we obtain that $H\Gamma _{1}^{0,0}$ consists of
countable number of components labelled by an integer. Let us fix one
operator $T_{k}$ in every component and define $\mathcal{C}_{m}^{k}$as a set
of gauge fields of the form (3) with $T=T_{k}$ and $\rho _{\mu }\in \Gamma
^{m}=\underset{m^{\prime }<m}{\cup }\Gamma _{1}^{m^{\prime }}$.
Every gauge field $\nabla _{\mu }\in \mathcal{C}_{m}$ is gauge equivalent to
a gauge field from $\mathcal{C}_{m}^{\prime }=\underset{k\in \mathbb{Z}}{%
\cup }\mathcal{C}_{m}^{k}$. (Recall that we consider the group of unitary
endomorphisms $U($Mat$_{n}(M(A)))$ as a gauge group.) We see that we can
restrict ourselves to the gauge fields from $\mathcal{C}_{m}^{\prime }$. 
The
gauge group becomes smaller after this restriction.

Similar statements are correct in any dimension. There are some
complications related to the fact that in general the group $\pi
_{d-1}(U(n)) $ does not coincide with stable homotopy group $\pi
_{d-1}(U(\infty ))=\pi _{d-1}(U_{\infty }(\mathbb{C})).$ However, in
noncommutative case ($\theta \neq 0$) the sets $\mathcal{C}_{m}^{k}$ and $%
\mathcal{C}_{m}^{l}$ defined by means $T_{k}$ and $T_{l}$ correspondingly
are related by gauge transformation if $T_{k}$ and $T_{l}$ determine the
same element of $\pi _{d-1}(U(\infty ))$. ( A proof for nondegenerate $%
\theta $ is given below$.$)This means that for odd $d$ we need only one
$T,$
and for even $d$ we should take $\mathcal{C}_{m}^{\prime }=\cup _{r\in 
\mathbb{Z}}\mathcal{C}_{m}^{r}$ where the index $r$ labels elements of $\pi
_{d-1}(U(\infty ))=\mathbb{Z}$.

It is easy to check that a unitary endomorphism $\varphi =1+\tau $, where $%
\tau $ is a matrix with entries from $\Gamma ^{m+1}$ transforms $\mathcal{C}%
_{m}^{\prime }$ into itself. It is convenient to consider the group $%
G=G^{(m+1)}$ consisting of endomorphisms of such a kind as residual gauge
group, that remains when we restrict ourselves to the gauge fields from $%
\mathcal{C}^{\prime }=\mathcal{C}_{m}^{\prime }$. (We omit the index $m$ in
topological considerations, because homotopy groups of $G$ and $\mathcal{C}%
^{\prime }$ don't depend on $m$.) It is easy to describe the topology of the
group $G$. If $\theta $ is nondegenerate then $\pi _{i}(G)=\pi
_{i}(U_{\infty }(\mathbb{C}))$; this homotopy group vanishes for odd $i$ and
is isomorphic to $\mathbb{Z}$ for even $i$ (Bott periodicity theorem). This
statement becomes almost obvious if we take into account that the group $G$
lies between $U_{\infty }(\mathbb{C})$ and the group $\mathcal{K}$ of
unitary transformations of the form $1+\tau $ where $\tau $ is a compact
operator. (See [7], [8],[9] for the analysis of topological properties of
different spaces of operators in infinite-dimensional case.)

If $\theta =0$ it is easy to check that $\pi _{i}(G)=\pi _{i+d}(U(n))$. If $%
\theta $ is degenerate, but $\theta \neq 0$ we have for even $d$ the same
answer as for the case of non-degenerate $\theta $; for odd $d$ we obtain
that $\pi _{i}(G)=\pi _{i+d+1}(U_{\infty }(\mathbb{C}))$ (i. e. $\pi
_{2k}(G)=0$, $\pi _{2k+1}(G)=\mathbb{Z}$). The calculations for degenerate $%
\theta $ are based on the remark that in this case an element of $G$ can be
considered as a map of $S^{d-\text{rank }\theta }$ into $\mathcal{K}$. (If $%
\theta $ is degenerate, but not equal to zero, we can assume that the first $%
d-$rank $\theta $ coordinates commute with all coordinates and the last rank 
$\theta $ coordinates obey canonical commutation relations. Considering the
first coordinates as parameters we obtain that an element of $G$ can be
regarded as a map of $\mathbb{R}^{d-\text{rank }\theta }$ into $\mathcal{K}$%
. This map can be extended to a continuous map of $S^{d-\text{rank }\theta }$
into $\mathcal{K}$.)

The group $G$ deserves the name of gauge group of Yang -Mills theory on $%
\mathbb{R}_{\theta }^{d}$ if we are working only with gauge fields from $%
\mathcal{C}^{\prime }=\mathcal{C}_{m}^{\prime }$.This means, in particular,
that corresponding functional integral can be taken over $\mathcal{C}%
^{\prime }/G$. (Notice that $\mathcal{C}^{\prime }$ is a disjoint union of
contractible sets, therefore it is easy to analyze the topology of $\mathcal{%
C}^{\prime }/G$ using the results above). However, one can show that the
group $\widetilde{G}$ consisting of unitary endomorphisms (of elements of $U(
$Mat $_{n}(M(A)))$ ) that transform $\mathcal{C}^{\prime }$ into itself is
larger then $G$. This follows from the consideration below, but it is
possible to show this directly. Namely, if the commutator  of  unitary
endomorphism $U$ with the operator $T_{k}$ that enters the definition of $%
\mathcal{C}^{\prime }$belongs to $\Gamma ^{m\text{ }}$then $U\subset $ $%
\widetilde{G}$. It is easy to construct examples of such endomorphisms that
don't belong to $G,$but it is not so easy to give a complete description of  
$\widetilde{G}$ . 

There exists another language that is more convenient to deal with fields
gauge trivial at infinity. Let us consider at first the case when the
parameter of noncommutativity $\theta $ is a non-degenerate matrix. Then the
dimension $d$ is even and the algebra $S(\mathbb{R}_{\theta }^{d})$ is
isomorphic to the algebra of integral operators acting on the space $S(%
\mathbb{R}^{p})$ where $2p=d$ and having a kernel, belonging to
$S(\mathbb{R}%
^{d})$. This means that we can consider $S(\mathbb{R}^{m})$ as a $A$-module;
we denote this module (Fock module) by $\mathcal{F}.$ The $\func{mod}$ule $%
\mathcal{F}$ can be considered as a Hilbert module over $A=S(\mathbb{R}%
_{\theta }^{d}).$ We assume that in the definition of gauge triviality at
infinity we have $T\in H\Gamma _{1}^{0,0}$ and $T^{+}T=1$ (i. e. $\Pi
^{\prime }=0$). Then one can prove that Ker$T^{+}=$Ker$TT^{+}=$Ker$%
(1-\Pi )=$Im$\Pi $ considered as $A$-$\func{mod}$ule is isomorphic to $%
\mathcal{F}^{r}$ for some $r\geq 0$. (The proof is based on a remark that $%
\Pi \in $ $S(\mathbb{R}_{\theta }^{d})$ obeys $\Pi ^{2}=\Pi ,\Pi =\Pi ^{+},$%
and therefore the corresponding integral operator is a projector on
finite-dimensional subspace of $S(\mathbb{R}^{p})$ .)

Using this fact we construct a map of $A^{n}$ onto $\mathcal{F}^{r}\oplus
A^{n}$ transforming $y\in A^{n}$ into $(\Pi y,T^{+}y)$. This map is an
isomorphism of $A$-modules. (The inverse map transforms $(\xi ,x)\in $Ker$%
T^{+}\oplus A^{n}$ into $\xi +Tx\in A^{n}$.)

Notice, that the isomorphism class of the module Ker $T^{+}$  
depends
only on the element of stable homotopy group $\pi _{d-1}(U(\infty ))$
determined by $T$. This remark proves the statement that gauge fields in $%
\mathcal{C}(T)$ are gauge equivalent to the fields in $\mathcal{C}%
(T^{^{\prime }})$ if $T$ and $T^{^{\prime }}$ determine the same element of $%
\pi _{d-1}(U(\infty ))$ (for the case when $\theta $ is non-degenerate).

Every connection on a module $\mathcal{F}_{rn}=\mathcal{F}^{r}\oplus A^{n}=$%
Ker $T^{+}\oplus A^{n}$ has the form 
\begin{equation}
\nabla _{\mu }=\nabla _{\mu }^{st}+\nu _{\mu }
\end{equation}%
where $\nabla _{\mu }^{st}$ stands for the standard connection $(i(\theta
^{-1})_{\alpha \beta }\widehat{x}^{\beta },\partial _{\alpha })=(\Pi
\partial _{\mu }\Pi ,\partial _{\mu })$ and%
\begin{equation}
\nu _{\mu }=\left( 
\begin{array}{cc}
M_{\mu } & N_{\mu } \\ 
R_{\mu } & S_{\mu }%
\end{array}%
\right)
\end{equation}%
is an endomorphism of $\mathcal{F}_{rn}$ represented by a block matrix where 
$M_{\mu }$ is an $r\times r$ matrix with entries from $\mathbb{C}$, $N_{\mu
} $ is an $r\times n$ matrix with entries from $\mathcal{F}$, $R_{\mu }$ is
an $n\times r$ matrix with entries from $\overline{\mathcal{F}}$, and $%
S_{\mu }$ is an $r\times r$ matrix with entries from $M(A)$.

Notice that in the above consideration instead of $A=S(\mathbb{R}_{\theta
}^{d})$ we can consider other algebras; for example, one can take $A=\Gamma
^{m}$ with $m\leq 0$.

Let us consider now a gauge field (3) where $\rho _{\mu }\in \Gamma ^{m}$
(i. e. a gauge field from the class $\mathcal{C}_{m}$). Then it is easy to
check that the for corresponding gauge field on $\mathcal{F}_{rn}$ we have $%
S_{\mu }\in \Gamma ^{m}$. (More precisely, $S_{\mu }$ has the same behavior
at infinity as $\rho _{\mu }$). This means that instead of gauge fields that
belong to the class $\mathcal{C}_{m}^{r}$ we can work with the class $%
\widetilde{\mathcal{C}}_{m}^{r}$ consisting of gauge fields on $\mathcal{F}%
_{rn}$ obeying $S_{\mu }\in \Gamma ^{m}$. (We constructed a one-to-one
correspondence between $\mathcal{C}_{m}^{r}$ and $\widetilde{\mathcal{C}}%
_{m}^{r}$.)

The gauge group  in this formalism should be considered as the group of
unitary endomorphisms of $\mathcal{F}_{rn}$ that are represented by matrices
of the form%
\begin{equation}
\left( 
\begin{array}{cc}
\mathcal{M} & \mathcal{N} \\ 
\mathcal{R} & \mathcal{S}%
\end{array}%
\right) 
\end{equation}%
where $\mathcal{S}-1\in \Gamma ^{m+1}$. Due to the correspondence between $%
\mathcal{C}_{m}^{r}$ and $\widetilde{\mathcal{C}}_{m}^{r}$ this group can be
considered also as a group of gauge transformations acting on $\mathcal{C}%
_{m}^{r}$ . Imposing conditions $\mathcal{M}=1$, $\mathcal{N}=\mathcal{R}=0$%
, we obtain transformations of $\mathcal{C}_{m}^{r}$ belonging to $%
G=G^{(m+1)}$; if these conditions are not satisfied we obtain gauge
transformations of $\mathcal{C}_{m}^{r}$ that don't belong to $G,$but belong
to $\widetilde{G}$. We see that the gauge group is larger than $G$.However,
it is easy to verify, that its topological properties are the same.They
coincide with the topological properties of the gauge group considered in
[1]. 

It seems that the simplest way to work with noncommutative gauge theories on 
$\mathbb{R}^{d}$ is to consider unitized algebras. (This is the viewpoint
advocated in [2].) We can reformulate the above consideration working with
the algebra $\Gamma ^{m}$ and corresponding unitized algebra $\widetilde{%
\Gamma }^{m}$. The gauge fields from $\widetilde{\mathcal{C}}_{m}^{r}$ are
precisely the fields that can be regarded as connections on $\mathcal{F}_{rn}
$, considered as $\widetilde{\Gamma }^{m}$-module.

Let us consider now the case when the dimension of the space
$(\mathbb{R}^{d})$ is less than 4. In this case  
 fields of the form
$\nabla_{\mu}=\partial _{\mu}+\rho _{\mu}$
where $\rho_{\mu}\in \Gamma ^m$ and $m=1-{d\over 2}$ have finite
euclidean action.Let us denote the class of fields of this kind
by $\cal D$.We expect that "almost
all'' gauge fields having finite euclidean action  are gauge equivalent
to the fields belonging to $\cal D$.It is easy to check that a unitary
endomorphism corresponding to a matrix  $T\in H\Gamma _{1}^{0,0}$,
transforms $\cal D$ into itself.The group consisting of endomorphisms 
of this kind be considered as a gauge group. 

{\bf Acknowledgments}

I am indebted to M. Kontsevich,  N. Nekrasov and M. Rieffel for useful
remarks. 
  
{\bf Appendix}.

Let us denote by $(\mathbb{R}^{d})$ the class of smooth matrix functions $%
a(z)$ on $\mathbb{R}^{d}$ obeying 
\begin{equation}
\left\| \partial _{\alpha }a(z)\right\| \leq C_{\alpha }\left\langle
z\right\rangle ^{m-\rho \left| \alpha \right| }
\end{equation}%
where $\alpha =(\alpha _{1},...,\alpha _{d})$, $\left| \alpha \right|
=\alpha _{1}+...+\alpha _{d},m\in \mathbb{R}$, $0<\rho \leq 1$, $%
\left\langle z\right\rangle =(1+\left\| z\right\| ^{2})^{1/2}$. We define
star-product of matrix functions using star-products of their entries and
standard rules of matrix multiplication. (The star-product $a\star _{\theta
}b$ as always depends on antisymmetric matrix $\theta $.) One can prove that
the star-product of functions $a^{\prime }\in \Gamma _{\rho }^{m_{1}}$ and $%
a^{^{\prime \prime }}\in $ $\Gamma _{\rho }^{m_{2}}$ belongs to $\Gamma
_{\rho }^{m_{1}+m_{2}}$. (In particular, $\Gamma _{\rho }^{m}$ is an algebra
if $m\leq 0$).

A matrix function $a(z)$ belongs to the class $\widetilde{\Gamma }_{\rho
}^{m}(\mathbb{R}^{d})$ if

\begin{equation}
\left\| a(z)\right\| \leq \text{const}\cdot \left\langle z\right\rangle
^{m},\left\| \partial _{\alpha }a(z)\right\| \leq C_{\alpha }\left\|
a(z)\right\| \left\langle z\right\rangle ^{-\rho \left| \alpha \right| },
\end{equation}%
(This condition is stronger than (7).).

One says that $a$ $\in H\Gamma _{\rho }^{m,m_{0}}$ if $a\in \widetilde{%
\Gamma }_{\rho }^{m}$ and $a^{-1}\in \widetilde{\Gamma }_{\rho }^{-m_{0}}$.
(We don't assume that $a(z)$ is invertible for all $z\in \mathbb{R}^{d}$,
however, we suppose that $a^{-1}(z)$exists for sufficiently large $\left\|
z\right\| $.) One can prove that for every function $a(z)$ from $H\Gamma
_{\rho }^{m,m_{0}}$ and for any $\theta $ in the definition of star-product
there exists such a function $b(z)\in \Gamma _{\rho }^{-m,-m_{0}}$ that $%
1-a\star _{\theta }b$ and $1-b\star _{\theta }a$ are matrix functions with
entries from $S(\mathbb{R}^{d}).$

\bigskip \textbf{References}

\bigskip 1. J. A. Harvey, \textit{Topology of the Gauge Group in
Noncommutative Gauge Theory,} hep-th/0105242

2. A. Schwarz, \textit{Noncommutative instantons: a new approach, }hep-th/
0102182

3. A.Connes, \textit{C*-algebres et Geometrie Differentielle, }C.
R. Acad. Sci.\textit{\ }Paris \textbf{290 }\textit{(1980), 599-604,
}English translation hep-th/0101093

4. A.Connes, \textit{Noncommutative Geometry, }Academic Press (1994), 1-655

5. A. Schwarz, \textit{Noncommutative supergeometry and duality, } Lett.
Math. Phys. \textbf{50} (1999), 309-321, hep-th/9912212

6. M. Shubin, \textit{Pseudodifferential operators, }Berlin, Springer
(1987), 1-278

7. A. Schwarz, \textit{On homotopic topology of Banach spaces, } Dokl. Akad.
Nauk SSSR, 154(1964), 61--63

8. R. S. Palais, \textit{\ On the homotopy type of certain groups of
operators, }Topology, (1964)

9. N. H. Kuiper, \textit{The homotopy type of the unitary group in Hilbert
space, }Topology, \textbf{3}, 1 (1965), 19-30.

10. K. K. Uhlenbeck, \textit{The Chern classes of Sobolev connections, }
Comm. Math. Phys., 101 (1985) 449-457.

11.M. Antonets, \textit{ Classical limit of Weyl quantization}
Teoret.Mat. Fiz. 38 (1979)331-344

12.J. M. Gracia-Bondia, J. Varilly, \textit {
Algebras od distributions suitable for phase-space quantum mechanics}
J. Math.Phys. 29 (1988) 869-879

13.R. Estrada, J. M. Gracia-Bondia, J. Varilly, \textit {
On asymptotic expansion of twisted products.} J. Math. Phys.
30 (1989) 2789-2796

\end{document}